# Web Services for the Virtual Observatory

Alexander S. Szalay, Johns Hopkins University
Tamás Budavári, Johns Hopkins University
Tanu Malik, Johns Hopkins University
Jim Gray, Microsoft Research
Ani Thakar, Johns Hopkins University

# Web Services for the Virtual Observatory

Alexander S. Szalay[a], Tamás Budavári[a], Tanu Malik[a,b], Jim Gray[b], and Ani Thakar[a]

[a]Department of Physics and Astronomy, The Johns Hopkins University, Baltimore, MD 21218
[b]Microsoft Research, San Francisco, CA 94105

## ABSTRACT

Web Services form a new, emerging paradigm to handle distributed access to resources over the Internet. There are platform independent standards (SOAP, WSDL), which make the developers' task considerably easier. This article discusses how web services could be used in the context of the Virtual Observatory. We envisage a multi-layer architecture, with interoperating services. A well-designed lower layer consisting of simple, standard services implemented by most data providers will go a long way towards establishing a modular architecture. More complex applications can be built upon this core layer. We present two prototype applications, the SdssCutout and the SkyQuery as examples of this layered architecture.

Keywords: Virtual observatory, web services, metadata, databases, world wide web

## 1. INTRODUCTION

### 1.1 Living in an exponential world

Astronomical data is growing at an exponential rate: it is doubling approximately every year. The main reason for this trend is Moore's Law, since our data collection hardware consists of computers and detectors based on the same technology as the CPUs and memory. However, it is interesting to note how the exponential trend emerges. Once a new instrument is built, the size of the detector is frozen and from that point on the data just keeps accumulating at a constant rate. Thus, the exponential growth arises from the continuous construction of new facilities with ever better detectors. New instruments emerge ever more frequently, so the growth of data is faster than just the Moore's Law prediction. Therefore while every instrument produces a steady data stream, there is an ever more complex network of facilities with large output data sets over the world[1].

How can we cope with this trend? First of all, since the basic pipelines processing and storage are linearly proportional to the amount of data, the same technology that gives us the larger detectors, will also give us the computers to process and the disks to save the data. On a per project basis the task will probably get easier and easier, the first year will be the most expensive, later it becomes increasingly trivial. On the community level however, the trend is not so clear, as we show below. More and more projects will engage in data intensive projects, and they will have to do much of the data archiving themselves. The integrated costs of hardware and storage over the community will probably increase as time goes on, but only slightly.

The software is where the difficulties are and will be. The software technology used in astronomy today has its roots in Fortran, with more and more C emerging, but object oriented technologies are relatively recent. Components are rarely reused. Projects tend to write their own software, using relatively few common libraries, and as a result the software costs are claiming a growing fraction of project budgets – 25% to 50%. This is especially true, if we look at the problem from a community-wide perspective.

### 1.2 Making discoveries

The strongest motivation for building new sky surveys is to make new discoveries. It is important therefore to consider when and where are new discoveries made. We believe that the new discoveries are almost always made at the edges or frontiers: either we need to look much deeper and detect fainter objects, or we have to go to extreme colors, by selecting the edges of a color distribution. We can search for objects of extreme shapes (gravitationally lensed arcs), or of extreme time-domain behavior (supernovae, microlensing).

When the Internet was in its infancy, Bob Metcalfe, an inventor of the Ethernet postulated Metcalfe's Law: *the utility of a computer network is proportional to the square of the number of nodes*. It is the number of the different connections one can make that matters! This law applies to discoveries as well: it is the number of connections we can make between fundamental properties that enable us to make new discoveries. A new observation of the sky in a previously unobserved wavelength or a new epoch for time-domain astronomy each enables new connections to be made. The utility of a collection of independent observations is proportional to the number of non-trivial connections among them. This is the motivation behind building multi-wavelength sky surveys. By federating multiple, independent projects, we can make these new connections. The early successes of today's sky surveys, SDSS and 2MASS have clearly proven this, the number of discoveries they made after the first few hundred square degrees of observations (high redshift quasars, brown dwarfs) were clearly out of proportion to the area of sky. They can only be explained when we include the possible number of pair-wise comparisons between filters (an amplification of connections by 5x4/2=10 for SDSS, 3x2/2=3 for 2MASS, 8x7/2=28 for their combination).

### 1.3 Publishing scientific data

It is generally believed that the scientific data publishing process is well understood. There are the *Authors*, mostly individuals or small groups, who create the experiments that provide the data, and they write papers that contain the data. There are the *Publishers*, the scientific journals, which print the papers, and nowadays also make them available in an on-line version, there are the *Curators*, whose role is filled by the libraries, which organize and store the journals, and make them available for the *Consumers*, who are scientists, who want to use and cite the data in their own research.

This model worked extremely well over the last hundred years, when all the scientific data relevant to the research could easily be included in the tables of the publication. This imbed-the-data-in-the paper publication model breaks down with the emergence of large datasets. We know about the large data sets in astronomy, there are even larger quantities of data in particle physics, and a similarly complex picture is emerging in genomic and biology research and in many other disciplines.

The above roles are clearly present in data intensive science, but they are performed in a very different fashion. The role of Authors belongs to large collaborations, like the SDSS, or the Human Genome Project, or CMS. They take five to ten years to build their experiment, before it starts producing data. The data volume is so large that it will never be contained in journals—at most small summaries or graphs will be printed. The data is made available to the collaborations (and the world) through web-based archives. During the project lifetime the curation responsibility rests with the projects themselves. Once the collaboration moves on to do something else, the data is either discarded, or moved to a national archive facility. The Consumers have to deal with the data in these many forms, often obtaining it from sources that are not eager to support them. The economic model for long-term curation is difficult since the costs fall to one group and the benefits to another.

### 1.4 Changing roles

The exponential growth in the data sources and the exponential growth in the individual data sets put a particular burden on the projects. It only makes sense to spend 6 years to build an instrument, if one is ready to use the instrument for at least the same amount of time. This means, that during the lifetime of a 6-year project, the data growing at a linear rate, the mean time the data spends in the project archive before moving to the centralized facility is about 3 years. Turning this around, the data that makes it into the national facilities will be typically 3 years old. As the amount of data is doubling in the world, every year, in 3 years the data grows by 8-fold, thus no central archive will contain more that about 12% of the world's data at any one time. The vast majority of the data and almost all the "current" data will be decentralized among the data sources. This is a direct consequence of the patterns of data intensive science. These numbers were of course taken from astronomy, the rates may be different for other areas of science, but the main conclusions remain the same.

This means that the projects are much more than just Authors: they are also Publishers, and to a large extent Curators. While scientists understand well authorship, they are less familiar with the responsibility of the other two roles. This is the reason why so many projects are spending large amounts of money on the software, and not just on the basic pipeline reductions. As many projects are experimenting with these roles, much effort is duplicated and much development wasted. We need to identify the common design patterns in the publishing and curation process, and to build reusable components for general availability.

## 2. WEB SERVICES: USING DISTRIBUTED DATA

**2.1 XML, SOAP, WSDL**

These problems are not unique to science: the same issues are emerging in the business world, where companies need to send and receive lots of diverse information not only inside their corporate firewalls, but also with the outside. Exchanging and automatically reading data in various formats have been hunting developers for many years. Finally, there is a world wide standard emerging for data representation in the form of XML, the Extensible Markup Language. There are clear grammatical rules for encapsulating complex information in a machine-readable form. XML is rather complex and it was not designed to be human readable. There are style sheets, which render the XML data to various easily understandable formats.

The most recent developments are related to Web Services: a standardized way to invoke remote resources on the web and exchange complex data. Indeed, web services define a distributed object model that lets us build Internet-scale software components and services. The major commercial vendors have agreed on SOAP (Simple Object Access Protocol) that specifies how to invoke applications that can talk to one another and exchange complex data. The Web Service Description Language (WSDL) enables an application to find out the precise calling convention of a remote resource, and build a compatible interface. Several toolkits, many of them freely available link web services to most of the modern programming languages and hardware platforms.

**2.2 Web Services in the Virtual Observatory**

Many of the expected tasks in the Virtual Observatory map extremely well to web services. Astronomers are already accustomed to various analysis packages, like IRAF, IDL or AIPS++, which have multi-layer APIs. They start with layer of simple image processing tasks. Then they build a layer of much more complex processing steps on top of this layer. These packages assume that the data resides in FITS files in the local file system, and the processing is done on the workstation itself.

With the Virtual Observatory[2] the main difference is that most the data will be remote. As a result, data access to remote resources needs to be just as transparent as if it were local. The remote data volume may be huge; therefore, it makes sense to move as much of the data processing as near the data as possible, because in many cases after the first few steps of processing the output volume is dramatically smaller (e.g. extracting object catalogs). In many cases the data will not only be remote, but it does not even exist at the time of the request: it may be extracted from a database with a query created at that moment. One can carry this even further, the requested data may be created by a complex pipeline on the fly, according to the user's specification, like a recalibration and custom object detection ran on an image built as a mosaic from its parts. These are called 'Virtual Data'—data that is created dynamically from its archived components.

**2.3 Everything for everybody?**

We believe that a multilevel hierarchy of services is the answer for the VO architecture. IRAF and AIPS++ are prototypes of this, but the concept needs to be extended to handle remote and virtual data sources. The core will be set of simple, low level services that are easy to implement even by small projects. Thus the threshold to join the VO will be low. Large data providers may be able to implement more complex, high-speed services as well. These core services can be combined into more complex portals that talk to several services, and create more complex results. Such a design will have modularity of components, standardization of interfaces, and access to commercially built toolkits for the lowest level communication tasks. We need to focus on the astronomy specific issues, and need not reinvent the wheel.

We need to choose carefully how we define the VO framework, and how many core applications we develop. The resources for development are scarce, and they need to be used efficiently. It is important to emphasize that it would be a mistake to try to build everything for everybody. It is impossible to make everybody happy. How to choose then?

There is a well-known rule about how many people are reading certain web pages. It says that 10% of the web pages are viewed by 90%. In a well-designed system there are always components which are much more popular than others. If we spend public money to build 10% of all possible applications we can think of, but these make 90% of the astrophysics community happy, we have satisfied our goals. There is no need to build all the possible complex applications—our colleagues have been very resourceful to build quite complex applications out of the IRAF modules. We need to provide clear standards, interfaces and documentation, have most data providers adopt a set of core services, build 10% of the system, then the rest of the custom applications will be built by individual astronomers. Efficient low-level services, and templates for the portals will save time for everybody. The details, whether we need a 90-10 or 80-20 rule, are irrelevant as long as we do not lose sight of our ultimate goal: to create a system that is simple to use.

## 3. THE HIERARCHY OF WEB SERVICES

### 3.1 Core services

These are a set of web services, which perform very simple tasks, and do not rely on complex connections. Mostly they sit on top of a single resource, or catalog. They deal with everyday tasks that most of the current astronomy archives already provide. It should be very easy to implement them, and there should be templates available for new resources.

- *Metadata services:* These tell about the particular archive or resource, what is its waveband, what is its sky coverage, what are the tables in the database, what are the fields in the tables, what are their physical units, their description.
- *Simple search patterns:* These will provide all the objects within a radius of a given point (cone-search[3]), or an image mosaic, assembled from parts, centered on a given point, in standard orientation and at a specified scale.
- *Simple functions:* Filtering on certain parameters, unit conversions, on-the-fly recalibrations, counts, simple histograms, plots belong to this category.

### 3.2 Higher level services

These are built on top of the atomic services, and perform more complex tasks. We expect to build such portals in a few days from the existing components and templates, like in IRAF today. Generally they talk to more than one service. Some examples include:

- *Automated resource discovery:* find the archives, which have optical and infrared data about a certain part of the sky, or find archives which have spectra at higher than 1 Angstrom resolution.
- *Cross-identification:* find overlaps between archives and search for matching objects, satisfying certain selection criteria, like brown dwarfs, or high redshift quasars.
- *Photometric redshifts:* using multicolor data one can estimate statistically the redshifts of galaxies, even if no distances were measured. These can be used to build rest-frame selected samples of galaxies.
- *Outlier detections:* find objects at the extremes of the color distribution, which do not fit the patterns of 'typical' objects, like stars or normal galaxies. This is a very common pattern, which could be automated, and maps well on a web service.
- *Visualization facilities:* given the complex multidimensional data, it is still best to look at it visually. The brain has better pattern recognition capabilities than most of the AI software. It is important to create easy to use interfaces to build visualizations of at least well-known patterns, like multidimensional scatter-plots or density maps.
- *Statistical services:* many statistical analysis tools could be ran remotely, either near a data source, or on a portal, after a federation step. Such a way the user does not have to worry about compiling and building these tools, they could work like a web-based translation engine.

# 4. EXAMPLE: SKYQUERY AND COMPONENTS

**4.1 SkyQuery functionality**

As a step towards understanding web services, we built a simple application based on a web service hierarchy. SkyQuery is a portal[4] that federates five geographically separate archives and web services. It performs fuzzy spatial joins between three databases, SDSS, FIRST and 2MASS and creates queries that filter on attribute differences across the archives. It was built by Tanu Malik, Tamas Budavari and Ani Thakar in about 6 weeks using the Microsoft .NET framework. Most of the application code was written either in C# or in ANSI C. The underlying databases used SQL databases running our spatial search library[5].

The main interface to the application is through a web page that accepts a query through a typical web form. The SkyQuery web service calls a SkyNode web service on each of the archives, to perform the atomic tasks, plus another low-level service, the image cutout, which delivers a picture of the search area.

```
SELECT o.objId, o.r, o.type, t.objId, t.m_j
  FROM SDSS:PhotoPrimary o, TWOMASS:PhotoPrimary t
 WHERE XMATCH(o,t)<3.5 AND AREA(181.3,-0.76,6.5)
       AND o.type=3 and (o.i - t.m_j)>2
```

**Figure 1.** The query syntax for the SkyQuery. There are special target designators to specify the archive, and there are two special operators, AREA and XMATCH, used to constrain the search area and the search accuracy.

SkyQuery is accessed by specifying an enhanced SQL query where the relational tables can be in different databases. It allows two kinds of clauses beyond regular SQL: (1) AREA to specify a spatial range; and (2) XMATCH to specify a probabilistic (fuzzy) spatial join. AREA is used to specify a region on the sky where we search for matches. This applies to all the databases referred to in the query. Currently we do not check whether the sky coverage of the database overlaps with the AREA. XMATCH is used to specify how to match objects in the different databases. The estimated position for the same object varies slightly from survey to survey as a normal random variable distributed around the real position. The standard deviation of the error in the measurement is assumed to be circular and is known for each survey. Given an object from each of the databases, one can compute the probability that these observations belong to the same common object. The parameter value of XMATCH specifies the threshold for such a set of objects.

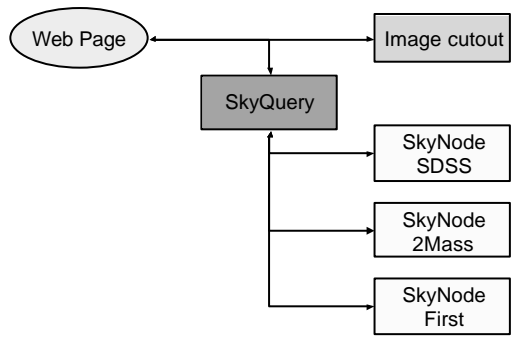

**Figure 2:** The relation of the different web services in the SkyQuery application. The SkyQuery web service calls the SkyNode web services on the participating archives. The ImageCutout provides an image display of the search area.

The XMATCH operator can be used to specify absence of a match as well. For example, if, instead, the clause was XMATCH(o,t,!p) < 3.5, it would say that return only those objects from the first two archives (o,t) that are within 3.5 standard deviations of a common position, and there is no matching object in the last archive (p) within the same error bound. The last archive is called a *dropout*. Archives specified in XMATCH are *mandatory*.

The heavy-duty cross-identification of the object lists is done iteratively at the SkyNodes. This enables very large savings in network transfer, since only the absolute minimal data is sent on the network. We estimated that our scheme requires about a factor of 50 less network transfers, when compared to a star-join topology that would stream all the data for a given area of the sky to the portal. Using this daisy-chain join, the portal can be very light weight, just sending questions and answers.

**4.2 The SkyQuery evaluation steps**

The SkyQuery portal follows this sequence of steps when it performs a cross identification of objects in the federation:

1. *Parse query*
2. *Get counts from federation members*
3. *Sort by counts*
4. *Make plan – smallest member first*
5. *Iterative cross match in smallest-member next daisy-chain*
6. *Select necessary attributes from last member*
7. *Return output*
8. *Return cutout image*

We built our own SQL parser to be able to handle the special operators. Once the parser identifies the different components, SkyQuery builds a query for each SkyNodes that counts the number of objects participating in the query. This query includes the region restriction specified by the AREA operator and the constraints that can be evaluated entirely at that node. These queries are a simple variant of a cone search, with some extra filters specified. The SkyQuery portal sorts count results by decreasing counts, and then creates a federated query plan. The query plan consists of a customized SQL query string for each of the archives that will be executed locally, after the cross-matching is done. This SQL SELECT statement will tell the archives, which attributes need to be sent on. The plan pipes the output of that node's query as input to the next node's spatial join.

In fact the architure pulls the data as follows. The portal calls the node with the largest count, trying to pull data from it. That node calls the next largest node, trying to pull data. Eventually the smallest (and last node is called, it is starts sending data up the pipeline. Results are returned as a DataSet objects. The returned dataset is converted to a temporary table by the web service, and then joined to its local database using the probabilistic cross-identification (see 4.3.4). This iteration eventually returns the final dataset to the SkyQuery service. The SkyQuery also uses the AREA descriptor for the parameters to the image cutout service, and returns the relevant optical sky image of the sky. In general a query involves were five interoperating web services (portal, SDSS[5], 2MASS[6], FIRST[7], CutOut[8]). Typical response times for a 10 arc minute radius are measured in seconds.

**4.3 The SkyNode**

]SkyNodes are the basic building blocks of the federated query system. They offer core services, including some sophisticated spatial search functions. This saves network transfer time. They are based on an identical web service: only the contents of the databases behind them differ. They support the following functions, described by the WSDL, also accessible though the SkyQuery web page.

4.3.1 Metadata functions

Each SkyNode is a stand-alone archive. The different functions of the SkyNode web service provide the low-level services mentioned above. These include metadata services:
- *Info():* returns a set of (key,value) pairs that specify the survey sky coverage, its wavelength coverage, and its root mean square circular positional accuracy.
- *Schema():* returns the XML schema of the whole archive, in the form of an empty DataSet document with the schema header.

- *Tables():* returns an XML document containing the list of tables accessible to the users with a short description of each table.
- *Columns(table):* lets the users query the database about the names, types, physical units and descriptions of the attributes. It returns an XML document containing a list of table.field and a short description of that field.
- *Functions():* returns the names of all functions and their parameters that the user can call, or use in queries
- *DocSearch(key):* lists any Table, Column or Function in the database that has the keyword either in its name, unit or description

These low-level metadata services enable the users (and other web services) to discover the metadata about the archive in a piece-by-piece fashion. Instead of dumping a large and complex document all at once, one can carry out a lazy evaluation: selectively retrieve only parts of the schema.

4.3.2 Query functions

There are two basic query functions, one is simple and the other is somewhat higher level. One could add a ConeSearch functionality, although through the use of the spatial search functions, Query() can fulfill this role as well. They are:
- *Query(string):* is a general purpose query interface, that allows any valid SQL statement to be submitted. It returns a DataSet object
- *XMatch(…):* is a more complex function, that performs a complex cross-identification. It receives a plan at its input, which contains instructions how to proceed. It calls another SkyNode recursively, it builds a temporary table in the database, performs the cross-identification, then joins the result with the main photometry table, then selects the necessary attributes. The updated select list is returned as a DataSet. If it is at the end of the calling chain, it runs a query on itself.

4.3.3 Database

The SkyNode database architecture was cloned form the SDSS SkyServer[6]. It is based upon Microsoft SQL Server 2000. It took a remarkably short time to port the external datasets to this format, about 1 day for the FIRST[7] archive and about 1.5 weeks for the 2MASS[8] archive. We built a template of the SkyServer, so that only the main photometry table had to be rebuilt, but all the spatial search functions were able to run without modifications. This shows that it is possible to build a database template that makes the data publication process much easier. The documentation for the schema (units, descriptions) was created in sync with the database schema and was inserted into the database automatically.

4.3.4 Fuzzy Spatial Join

The algorithm used by a SkyNode to evaluate a cross match query is interesting. The AREA clause in a cross match query is implemented using the spatial search capabilities of the SkyNodes. We have built a library for spatial searches as an extended stored procedure to SQL Server, using the *Hierarchical Triangular Mesh* (HTM: Kunszt et al 2001). The library builds a quadtree on the sky, each node corresponding to a spherical triangle. For each object in the database we compute its 'trixel' address, or htmId, down to 20 levels on the quadtree. We can retrieve objects inside arbitrary spherical polygons. First we return the indices of the triangles that cover the area then we perform a join with the object table, using the clustered index built on the htmId field of the objects. The simplest search is the inside of a circle, centered on a point. Currently this is the only AREA specification that we support, but we will soon extend SkyQuery to more complex area definitions.

The XMATCH clause is implemented by computing a probabilistic spatial join. The log likelihood that the given objects in the different archives correspond to the same common object is obtained by minimizing the chi-square from the sum of squared deviations in the positions, using inverse variance weighting. We were using Cartesian coordinates $(x,y,z)$ forming a unit vector to describe a direction on the sky. This avoids any coordinate singularities, common in spherical coordinate systems. We compute a cumulative weighted direction $(a_x,a_y,a_z)$ and a cumulative weight ($a$) where the sum is taken over the archives for a given set of objects, a *tuple*. The best position is along the direction of $(a_x,a_y,a_z)$, and its log likelihood at that position can be derived from the cumulative quantities. For details see the documentation on the web site.

If there is more than one match to a given *(n-1)*-tuple in the next archive, we add new tuples as necessary. If there are no matches, the tuple is deleted. The tuples that survive until the end are returned as our final result. This cross-matching scheme is fully symmetric, the particular match order does not matter, other than in terms of speed.

**4.3 The Cutout service**

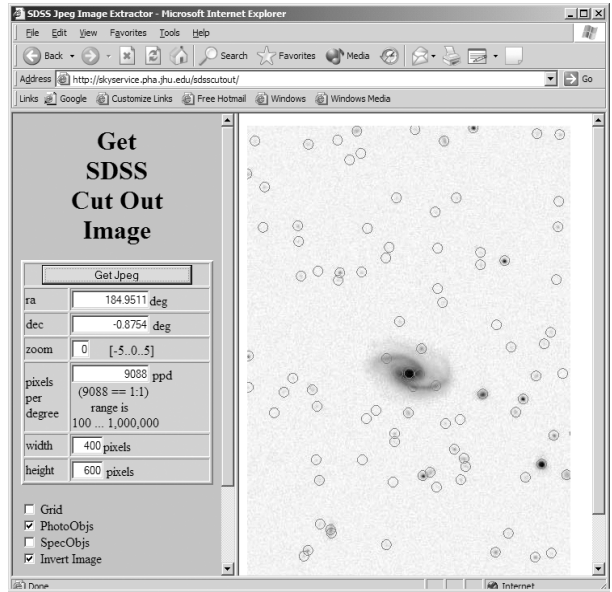

**Figure 3.** A screen shot of the web interface calling the cutout service. One can specify a location of the sky, a pixel scale (pixels per degree), and the size of the image in pixels.

We created a simple cutout service based on the images in the SDSS EDR archive, stored on the SkyServer. A web driver for the service is available at http://skyservice.pha.jhu.edu//sdsscutout/[9]. The full documentation is there as well. The image is created on the server dynamically, from several smaller images stored in the database. Their proper offsets are computed then they are shifted by the appropriate amount to the nearest pixel. No sub-pixel interpolation or image warping is used, nor is any necessary. The images are inserted to a main image buffer, which is then rotated, then the requested rectangle is cut out. It is rendered rotated, so that North is up, scaled to the desired pixels per degree, and one can plot optionally the inverse of the image. One can also overlay objects detected by the SDSS, or the ones that have measured spectra. It was implemented in C#. It took about a week of effort, and approximately a 500 lines of code to build this application. We were able to port a considerable amount of SkyServer's Javascript geometry code to C# without much trouble. The graphics tools of the .NET Framework Class Library enable us to do remarkably smooth image zooms over many orders of magnitude of scale factors. The cutout service has a dynamic range of 10,000 in image scaling.

## SUMMARY

In a world where data sizes and complexity are growing exponentially, we face new challenges. The usual paradigm of authoring and publishing data is changing, not to mention how we analyze the data. Many new discoveries require federating as many observations as possible. Due to the rapidly changing technology, these data sets are in distant corners of the world. In such a scenario there is no chance of building a centralized system. The large volume of data requires us to move the processing to the data, rather than the traditional way of moving the data to where the processing is.

Building software for more and more projects is a main challenge. For the community the only way to deal with this is to use a modular architecture that encourages code reuse. Also, the data publishing patterns are very similar and they yield well to the use of templates and design patterns. The recently introduced web services do most of this automatically.

We envisage a multiplayer network of interoperating web services that will form the foundation of the emerging Virtual Observatory. A well-defined set of core services implemented by most of the data providers will ensure a similarity in the basic functions and enable their automated integration at higher levels. If anything, it should be easier to build a new web service, than to link together several IRAF modules into a new application today.

As a test of these ideas we have implemented a two-layer application based entirely on web services. The SkyQuery, built on SkyNodes and a Cutout service has lived up to the promise, in about six weeks we were able to build a remarkably complex and flexible data federation service, which is only a small precursor to the greater and better things to come.

## ACKNOWLEDGEMENTS

AS acknowledges support from grants NSF AST-9802 980, NSF KDI/PHY-9980044, NASA LTSA NAG-53503 and NASA NAG-58590. He is also supported by a grant from Microsoft Research. Our effort has also been generously supported with hardware by Compaq, Intel and HP.